\magnification=\magstep1 \parskip=0pt
\fontdimen16\tensy=2.7pt\fontdimen17\tensy=2.7pt
\vsize=22truecm\hsize=16.5truecm 
\font\eightrm=cmr8  \font\eighti=cmmi8
\font\eightsy=cmsy8 \font\eightit=cmti8
\font\eightsl=cmsl8 \font\eightbf=cmbx8
\font\eighttt=cmtt8 
\def\eightpoint{%
\textfont0=\eightrm \scriptfont0=\fiverm \def\rm{\fam0\eightrm}%
\textfont1=\eighti  \scriptfont1=\fivei  \def\mit{\fam1}%
\textfont2=\eightsy \scriptfont2=\fivesy \def\cal{\fam2}%
\textfont\itfam=\eightit\def\it{\fam\itfam\eightit}%
\textfont\slfam=\eightsl\def\sl{\fam\slfam\eightsl}%
\textfont\bffam=\eightbf\def\bf{\fam\bffam\eightbf}%
\textfont\ttfam=\eighttt\def\tt{\fam\ttfam\eighttt}\rm}
\def\foot#1#2{{\baselineskip=10pt\footnote{$^{#1}$}{\eightpoint#2}}}
\def\dd{{\rm d}}\def\ee{{\rm e}}\def\ts{\textstyle}
\def\half{{\ts{1\over2}}}\def\de{\partial}
\def\section#1#2{\goodbreak \vskip0.8cm \leftskip=20pt \parindent=-20pt
  \indent{\bf\hbox to 20pt{#1\hss}#2}
  \bigskip\nobreak \leftskip=0pt \parindent=20pt}
\font\sf=msbm10\def\Bbb#1{\hbox{\sf #1}}\def\IR{\Bbb R}\def\ZZ{\Bbb Z}
\def\leftrightarrowfill{$\mathsurround=0pt \mathord\leftarrow \mkern-6mu
  \cleaders\hbox{$\mkern-2mu \mathord- \mkern-2mu$}\hfill
  \mkern-6mu \mathord\rightarrow$}
\def\overleftrightarrow#1{\vbox{\ialign{##\crcr
  \leftrightarrowfill\crcr\noalign{\kern-1pt\nointerlineskip}
  $\hfil\displaystyle{#1}\hfil$\crcr}}}
\def\nab{\nabla\!}\def\nabp{\nabla'\hskip-4pt}\def\tw{\tilde}
\def\dnabla{\overleftrightarrow{\nabla}\hskip-4pt}
\def\lapl{\kern0.5pt{\lower0.1pt\vbox{\hrule height.5pt width 6.8pt
    \hbox{\vrule width.5pt height6pt \kern6pt \vrule width.3pt}
    \hrule height.3pt width 6.8pt} }\kern1.5pt}


\line{\hfil RGGR-94/1}
\line{\hfil 22 February 1994}
\vfill
\centerline{\bf Relationships between various characterisations of wave
tails}
\vskip1truecm
\centerline{Luca Bombelli$^{\dag}$ and Sebastiano Sonego$^{\ddag}$}
\vskip.8truecm
\centerline{{\sl RGGR, Universit\'e Libre de Bruxelles,}}
\centerline{\sl Campus Plaine CP 231, 1050 Brussels, Belgium}
\vskip.3truecm
\centerline{$^{\dag}$ E-mail: lbombell@ulb.ac.be (Current e-mail:
luca@phy.olemiss.edu)}
\centerline{$^{\ddag}$ E-mail: ssonego@ulb.ac.be (Current e-mail:
Sebastiano.Sonego@Dic.Uniud.It)}
\vskip3truecm
\midinsert\narrower
{\baselineskip=8pt \eightpoint \centerline{Abstract} \medskip
\noindent One can define several properties of wave
equations that correspond to the absence of tails in their solutions, the
most common one by far being Huygens' principle. Not all of these
definitions are equivalent, although they are sometimes assumed to be. We
analyse this issue in detail for linear scalar waves, establishing some
relationships between the various properties. Huygens' principle is
almost always equivalent to the characteristic propagation property, and
in two spacetime dimensions the latter is equivalent to the zeroth order
progressing wave propagation property. Higher order progressing waves in
general do have tails, and do not seem to admit a simple physical
characterisation, but they are nevertheless useful because of their close
association with exactly solvable two-dimensional equations.}
\endinsert
\vfill
\noindent PACS numbers: 03.40.Kf; 02.30.Jr.\hfill\break
Short title: Characterisations of wave tails.
\eject


\section{1.}{Introduction}

\noindent The question whether a propagating wave leaves a tail behind it 
is one that has interested many physicists and mathematicians for a long
time, and has found applications from the first studies in light 
propagation [1] to the theory behind the proposed experiments
to detect gravitational waves [2,3]. Intuitively,
a wave tail can be described as follows. Suppose that, at some time $t$, an
instantaneous pulse of a field $\phi$ is produced at a point labelled by
spatial coordinates $\{x^i\}$. This pulse originates a wave front
propagating outward with some speed, that will be detected by an observer
sitting at some point $\{x'^i\}$ at a time $t'>t$. We say that the wave
has a {\it tail} or {\it wake} if such an observer continues to detect a
nonvanishing field even after the wave front has passed, i.e., at times
greater than $t'$. In this case we speak of diffusive, as opposed to
sharp, propagation.

One can roughly identify three distinct possible reasons for the occurrence
of tails:
\item{(i)} Mass-like terms in the wave equation. For example, the
Klein-Gordon equation
  $$
  \lapl\phi-\mu^2\phi=0
  \eqno(1.1)
  $$
in four-dimensional Minkowski spacetime has tails when $\mu\neq0$, but not
when $\mu=0$. This feature can be seen as the wave-mechanical counterpart
of the fact that massive particles move slower than the speed of light.
\item{(ii)} Dimensionality of spacetime. For example, the massless
Klein-Gordon equation in $m$-dimensional Minkowski spacetime has tails for
odd values of $m$, but not when $m$ is even, with the exception of $m=2$
[4--7].
\item{(iii)} Backscattering off potentials and/or spacetime curvature
[3,8,9]. This is clearly the most interesting tail production mechanism from
a physical viewpoint.
\vskip.2truecm

The study of the occurrence of tails, as well as of their implications, is
certainly of great interest and importance, but relies on the possibility
of using precise definitions. The heuristic characterisation given
above is evidently too loose for this purpose, but it can be easily
formalised to obtain a tail-free property. For linear wave equations,
however, a no-tails condition is better expressed in terms of the Green
function, and corresponds to what is usually called the Huygens
principle. Furthermore, still other definitions (e.g., characteristic
propagation property, progressing waves propagation property) can be found
in the literature. These characterisations are not all equivalent,
although they are sometimes (explicitly or implicitly) assumed to be. It
is the purpose of this paper to explore this issue in detail, by
establishing the relationships between them and clarifying their meaning
and domain of applicability.

If the wave equation is inhomogeneous (which corresponds, physically, to
the presence of sources), the formulation of no-tails properties is
a delicate task, because of the need to capture the notion of sharp
propagation when the field due to the sources is superposed to the one
arising purely from the boundary conditions. The situation is even worse
for nonlinear waves. For these reasons, we shall restrict ourselves
here to considering an arbitrary linear, second order homogeneous
hyperbolic partial differential equation, which can always be written in
the form
  $$
  g^{\mu\nu}\de_\mu\de_\nu\phi + h^\mu\de_\mu\phi + K\,\phi = 0\;,
  \eqno(1.2)
  $$
where $g^{\mu\nu}$, $h^\mu$, and $K$ are suitable real functions of $m$
variables $\{x^\mu\}$, that we can think of as coordinates over an
$m$-dimensional differentiable manifold $\cal M$; without loss of
generality, we can assume that $g^{\mu\nu}=g^{\nu\mu}$ so that these
functions can be thought of as the components of the inverse of a Lorentzian
metric $g_{ab}$ on $\cal M$ given, in the coordinates $\{x^\mu\}$, by the
inverse of the matrix $g^{\mu\nu}$; this metric, however, may not be the
one normally used to determine intervals and causal relations (for a physical
example in which it is not, see Ref 10). The metric $g_{ab}$ defines then a
Riemannian connection $\nabla$ that allows us to rewrite (1.2) equivalently as
  $$
  g^{ab}\nab_a\nab_b\phi + H^a\nab_a\phi + K\phi = 0\;, \eqno(1.3)
  $$
where $H^a$ is uniquely determined by $h^\mu$ and $g^{\mu\nu}$
(specifically, $H^\mu=h^\mu-\de_\nu g^{\mu\nu}-\de^\mu\ln\sqrt{-g}$, 
where $g$ is the determinant of $g_{\mu\nu}$).

Although physically one may wish to discuss the appearance of tails in
waves emitted by very distant sources, the phenomenon is essentially a
local one. Therefore, it will not represent a loss in generality to study
wave propagation close to the hypersurfaces where data are given, and we can
assume that $\cal M$ is globally hyperbolic and normal. A more general
manifold can always be covered by a collection of regions with these
properties, and tails will develop iff they already do inside some of
those smaller domains [11]. In such a manifold, a very powerful tool for
studying general properties of linear differential equations and their
solutions is that of Green functions [6,12]. A Green function $G(x,x')$
for (1.3) is defined by
  $$
  \Bigl(g^{ab}\nab_a\nab_b - H^a\nab_a + K-\nab_aH^a\Bigr)\,
  G(x,x') =-\delta(x,x')\;,
  \eqno(1.4)
  $$ 
and by appropriate boundary conditions; in (1.4), $\delta(x,x')$ is
the delta function on $\cal M$, such that for each test function $f$,
  $$
  \int\dd^mx'\,\sqrt{-g(x')}\,f(x')\,\delta(x,x') = f(x)\;. \eqno(1.5)
  $$ 
It should be noted that the differential operator acting on the  first
variable in $G$ is the adjoint of the one which acts on $\phi$ in (1.3),
and differs from it when $H^a\neq0$ [6].

If we choose to study the behaviour of the field at points to the future of
a Cauchy hypersurface $\cal S$ on which data are assigned, it is convenient
to work with the advanced Green function, which satisfies $G(x,x')=0$ when
$x'$ is to the past of $x$. (Of course, the whole discussion can be
carried out as well  for the ``time-reversed'' situation.)  In general,
since the wave equation gives rise to causal propagation ([13]; see also Ref
14, p 250), the advanced Green function can be written as
  $$
  G(x,x') = \tw\Sigma(x,x') + \tw \Delta(x,x')\;, \eqno(1.6)
  $$  
where $\tw\Sigma$ is a distribution term with support on light-like
separated pairs $(x,x')$, and $\tw \Delta$ has support on timelike related
ones, with $x$ to the past of $x'$ in both cases (in fact, this is a
possible {\it definition\/} of causal propagation). Thus,
$\tw\Delta$ is of the form
  $$
  \tw \Delta(x,x') =: \Delta(x,x')\,\theta_+(-\Gamma(x,x'))\;, \eqno(1.7)
  $$  
where $\Gamma(x,x')$ is the square of the proper distance calculated
along the unique geodesic connecting $x$ and $x'$, and $\theta_+$, with a
common abuse of notation, represents an ``advanced step function,'' which
is zero when $x$ lies in the future of $x'$ [6,12].

We will allow the Cauchy hypersurface $\cal S$ to be piecewise smooth,
like, e.g., the future null cone of a point, which is not differentiable
at the vertex. The field $\phi$ at each $x\in D^+({\cal S})\equiv
J^+({\cal S})$ can then be expressed as
  $$
  \phi(x) = -\int_{\cal S}\dd S_a(x')
  \Bigl[ g^{ab}(x')\,G(x',x)\dnabla_b\!'\phi(x')
  + H^a(x')\,G(x',x)\,\phi(x') \Bigr], \eqno(1.8)
  $$
where $\dd S_a(x')$ is the oriented volume element on the hypersurface
$\cal S$ at $x'$; this expression is valid even if $\cal S$ is a
(partially) null hypersurface, provided that $\dd S_a$ is appropriately
defined (see section 5).

We start in section 2 with a general discussion of the two-dimensional
case. In section 3 we give some definitions of properties of wave
propagation related to the absence of tails, whose meanings and mutual
relationships are investigated in sections 4--7. Section 8 contains some
concluding remarks, as well as speculations about possible lines of future
research on the topic. As we have already been doing, we will use latin
indices $a$, $b$, ... as abstract indices in spacetime [14], which just
indicate the tensorial nature of an object without requiring a set of
coordinates, while greek indices $\mu$, $\nu$, ... will be used for equations
valid in some chart. The notations $D^\pm(\cal A)$, $J^\pm(\cal A)$,
$I^\pm(\cal A)$, where $\cal A$ is some subset of $\cal M$, stand,
respectively, for its future/past domain of dependence, causal future/past,
and chronological future/past [14], all defined in terms of the causal
relations induced by $g_{ab}$.  Minkowskian coordinates, in which the
coefficients of the metric have (at least at a point) the form
$\eta_{\mu\nu}=\eta^{\mu\nu}={\rm diag} (-1,1,\ldots,1)$, will be denoted by
$(t,x^i)$, with $i=1,\ldots,m-1$.


\goodbreak
\section{2.}{The two-dimensional wave equation}

\noindent We begin with a general discussion of the wave equation (1.3)
in a two-dimensional spacetime. The motivation for devoting an entire
section to such a specific subject is twofold. First, it will provide us
with a concrete example with which to illustrate some general ideas. 
Second, and more important, much of the paper will turn out to deal
exclusively with this case; it is thus convenient to establish a few
preliminary points, to which we can refer later on.

Any two-dimensional spacetime is conformally flat, and its metric can
therefore be locally written as $g_{ab} = \Omega^2\eta_{ab}$, 
where $\Omega$ is a nonvanishing function and $\eta_{ab}$ the Minkowski
metric. In a general $m$-dimensional conformally flat spacetime we have
  $$
  \nab_aX^a=\de_aX^a+\de_a\ln|\Omega|^m\,X^a=|\Omega|^{-m}\de_a
  \left(|\Omega|^mX^a\right)\;,
  \eqno(2.1)
  $$
for an arbitrary vector field $X^a$. In the particular case $m=2$, and for
$X^a=g^{ab}\nab_bf=\Omega^{-2}\eta^{ab}\de_bf$, equation (2.1) allows us to
write
  $$
  g^{ab}\nab_a\nab_bf=\Omega^{-2}\eta^{ab}\de_a\de_bf\;,
  \eqno(2.2)
  $$
for any function $f$. Therefore, in $1+1$ dimensions,
  $$
  g^{ab}\nab_a\nab_bf+H^a\nab_af+Kf=\Omega^{-2}\Bigl(
  \eta^{ab}\de_a\de_bf+\bar{H}^a\de_af+\bar{K}f\Bigr)\;,
  \eqno(2.3)
  $$
where $\bar{H}^a:=\Omega^2H^a$ and $\bar{K}:=\Omega^2K$. Since
$\Omega\neq0$ everywhere on $\cal M$, we have that $\phi$ satisfies the
wave equation (1.3) in $({\cal M},g)$ iff it satisfies
  $$
  \eta^{ab}\de_a\de_b\phi+\bar{H}^a\de_a\phi+\bar{K}\phi=0
  \eqno(2.4)
  $$
in the flat spacetime $({\cal M},\eta)$. We can therefore restrict
ourselves to studying (2.4) without loss of generality.

It is convenient to introduce locally on $\cal M$ the null coordinates
  $$\eqalignno{
  &u:=\half\,(t-x)\;, &(2.5) \cr
  &v:=\half\,(t+x)\;, &(2.6) }
  $$
in which the Minkowski metric and its inverse have only one
independent nonvanishing component each, $\eta_{uv}=-2$ and
$\eta^{uv}=-1/2$, respectively, so that (2.4) becomes
  $$
  \de^2_{uv}\phi+U\,\de_u\phi+V\,\de_v\phi+W\phi=0\;,
  \eqno(2.7)
  $$ 
with $U:=-\bar{H}^u$, $V:=-\bar{H}^v$, and $W:=-\bar{K}$.
In the rest of the paper, we shall often refer to this form of the
two-dimensional wave equation. 

It is not difficult to check, using (2.1) again, that for an arbitrary
function $f$ in two dimensions,
  $$
  g^{ab}\nab_a\nab_bf-\nab_a\left(H^af\right)+Kf=
  -\Omega^{-2}\Bigl[\de^2_{uv}f-\de_u\left(Uf\right)-
  \de_v\left(Vf\right)+Wf\Bigr]\;.
  \eqno(2.8)
  $$
Since $\delta(u,v;u',v')=\Omega^{-2}\,\delta(u-u')\,\delta(v-v')$, it
follows that $G(u,v;u',v')$ is a Green function for (1.3) iff it
satisfies the equation
  $$
  \de^2_{uv}G-\de_u\left(UG\right)-\de_v\left(VG\right)+WG=
  \delta(u-u')\,\delta(v-v')\;.
  \eqno(2.9)
  $$
By direct substitution, we can verify that
  $$
  G(u,v;u',v')=\Delta(u,v;u',v')\,\theta(u'-u)\,\theta(v'-v)\;,
  \eqno(2.10)
  $$
where $\theta$ is the step function, satisfies (2.9) provided that the
following conditions are fulfilled:
  $$
  \de^2_{uv}\Delta-\de_u\left(U\Delta\right)-
  \de_v\left(V\Delta\right)+W\Delta=0\;;
  \eqno(2.11)
  $$
  $$
  \Delta(u,v';u',v')=\exp\left(-\int_u^{u'}\!\dd u''\,V(u'',v')\right);
  \eqno(2.12)
  $$
  $$
  \Delta(u',v;u',v')=\exp\left(-\int_v^{v'}\!\dd v''\,U(u',v'')\right).
  \eqno(2.13)
  $$
Since (2.12) and (2.13) can be regarded as data in a characteristic initial
value problem for (2.11), which has a unique solution, it follows that 
$\Delta(u,v;u',v')$ is completely determined by (2.11)--(2.13); therefore, 
(2.10) is the general form of the advanced Green function for the
two-dimensional wave equation. In the particular case $U=V=K=0$, we
recover the well-known result $\Delta=1$ [5,7].


\goodbreak
\section{3.}{Definitions}

\noindent Various properties are used to characterise the absence of wave
tails. First of all, we need to distinguish between the fact that a
particular solution of the wave equation may not have tails, and a possible
intrinsically non-diffusive nature of the equation itself. We say that a
set of Cauchy data with compact support $\cal C$ {\it produces no tails\/}
iff the field $\phi$ obtained evolving these data vanishes in $K^+({\cal C})$,
the set of all points in the causal future of $\cal C$ which cannot be
reached from $\cal C$ by a null geodesic [15]. For such data and points $x\in
K^+({\cal C})$, (1.8) becomes
  $$
  \phi(x) = -\int_{\cal C}\dd S_a(x')
  \Bigl[ g^{ab}(x')\,\Delta(x',x)\dnabla_b\!'\phi(x') +
  H^a(x')\,\Delta(x',x)\,\phi(x')
  \Bigr]\,\theta(-\Gamma(x',x))\;, \eqno(3.1)
  $$  
since the $\tw\Sigma$ term and the derivative of the $\theta$-function
in $G(x',x)$ do not contribute to the field at $x$ (we retain the factor
$\theta(-\Gamma)$ in order to cover also cases in which $x\in K^+({\cal
C})$ is not in the causal future of some $x'\in\cal C$, as it may happen, for
example, when $\cal C$ is not simply connected). Therefore, the data on
$\cal C$ produce no tails iff the integral on the right hand side of (3.1)
vanishes.

A related notion, expressed directly in terms of the field rather than of 
the Cauchy data, generalises the concept of distortionless propagation,
without necessarily referring to a differential equation [5]. We say that a
wave {\it propagates without distortion\/} if it can be written as $\phi(x) =
R(u(x))$, where the {\it phase\/} $u$ is a function on $\cal M$, and the {\it
wave form\/} $R$ is a function of one variable. Clearly, if $R$ has compact
support $[u_1,u_2]$, the wave propagates permanently sandwiched between those
two values of $u$. We have here no metric with which to verify whether this
wave has tails in the sense defined above, but it is nevertheless obvious
that ``there are no tails with respect to $u$.''  A slightly more general
situation is that of {\it relatively undistorted\/} or {\it simple
progressing\/} waves, those of the form $\phi(x)=f(x)\,R(u(x))$, where the
{\it amplitude\/} $f$ is a function on $\cal M$ [5,6,16--19]. An example in
Minkowski spacetime is the spherical wave $\phi(t,{\bf x})=\exp{\rm i}(k|{\bf
x}|-\omega t)/|{\bf x}|$, which progresses with speed $\omega/k$ in the
radial direction, and is undistorted except for the decrease in the amplitude
$1/|{\bf x}|$.\foot{\dagger}{Progressing waves become interesting for
suitable choices of the phase; the most common ones in Minkowski spacetime
have $u(t,{\bf x})=t-F({\bf x})$, for some wave front $F$, function of the
spatial coordinates [16].} A finite sum of simple progressing waves
will also be called (relatively) undistorted, the simplest example being the
general solution $\phi(t,x) = R(t-x) + S(t+x)$ of the  two-dimensional
equation (2.7) with $U=V=W=0$. A useful further generalisation is that of
{\it progressing waves of order\/}
$N$, those that can be written as
  $$
  \phi(x) = \sum_{i=0}^N f_i(x)\,R^{(i)}(u(x))\;, \eqno(3.2)
  $$
where the superscript $(i)$ denotes the $i^{\rm th}$ derivative. For
these waves it is still true that a compact support for $R$ implies a
``sandwich propagation'' for $\phi$.

Turning now to intrinsic properties of the wave equation, corresponding
to the absence of tails in the general solution, the most frequently used
one by far is Huygens' principle:\foot{\dagger}{The improper use of the
term ``principle'' to denote what is actually only a property is commonly
adopted for historical reasons, with the probably unique exception of
Hadamard, who refers to it as ``Huygens' minor premise'' [4]. In reality,
Huygens made use of TF below rather than HP, and merely in an {\it
approximate} version, by postulating that {\it almost} all the waves emitted
by a pointlike source are concentrated on the wavefront (see Ref 1, e.g.,
pp.\ 18 and 22). It is interesting to notice that this hypothesis does not
correspond to assuming that tails are absent, but only that they are small!}

\vskip2pt
\leftskip=20pt
\noindent{\bf HP:} A wave equation is said to satisfy the {\it Huygens
principle\/} (HP) iff the field at a point $x$ depends only on the data at
the intersection of the past light cone through $x$ with the Cauchy
hypersurface [4--6], in the sense that any two sets of data which coincide
there must give the same field at $x$.

\leftskip=0pt
\vskip3pt
\noindent It is obvious from (1.8) that HP is equivalent to the
requirement that the advanced Green function $G(x,x')$ have support only on
pairs of points such that $x$ lies on the past  light cone of $x'$. The
latter statement is often also referred to as ``Huygens' principle''. An
obvious consequence of (2.10) is that HP is always violated in a
two-dimensional spacetime.

The concepts discussed at the beginning of this section give rise, however,
to other definitions that can be found in the literature. Using the notion
of data that produce no tails, we define: 

\vskip2pt
\leftskip=20pt
\noindent{\bf TF:} A wave equation is said to be {\it tail-free\/} (TF)
iff each set of Cauchy data with compact support produces no tails [15].

\leftskip=0pt
\vskip3pt

\noindent It is almost trivial to see that TF and
HP are equivalent; nevertheless, for the sake of clarity, we shall present
the explicit proof in the next section.

It is sometimes interesting to consider data assigned on a characteristic
(i.e., null [6]) hypersurface, and to give therefore a modified version of TF.
This property was originally formulated in terms of solutions of a wave
equation [20], but we find it more meaningful to state it as applied to the
equation itself [15,21]:
\vskip2pt
\leftskip=20pt
\noindent{\bf CPP:} A wave equation is said to satisfy the {\it
characteristic propagation property\/} (CPP) iff each set of null data of
compact support, which vanish at any singular points of the support,
produces no tails.

\leftskip=0pt
\vskip3pt 
\noindent Why restrict, in this definition, the admissible data to those
that vanish at singular points?  For a two-dimensional characteristic
initial value problem, we can consider without loss of generality the
initial data $\phi(u,0) =:\varphi(u)$ and $\phi(0,v) =:\psi(v)$ to be
assigned on the union of the two ``half-axes'' $u\ge0$ and $v\ge0$. 
Consider the simplest example of (2.7), with $U=V=W=0$. Its general
solution is of the form $\phi(u,v) = R(u)+S(v)$, which in terms of the data
becomes $\phi(u,v)=\varphi(u)+\psi(v)-\phi(0,0)$. This means that, if
$\phi(0,0)\ne0$, no data with compact supports $[u_1,u_2]$ and $[v_1,v_2]$
can lead to ``sandwich propagation,'' i.e., one for which the support of
the full solution is contained in the union of the strips $u_1\le u\le u_2$
and $v_1\le v\le v_2$. There is, of course, nothing bad in this, since
we know already that HP is violated in two dimensions. However, data
$\varphi(u)$ and $\psi(v)$ such that $\varphi(0)=\psi(0)=\phi(0,0)=0$ do
lead to sandwich propagation. The point $(0,0)$ here is an example of
singular point of $\cal S$, and the reason we excluded data which do not
vanish at such points in the definition of CPP is that this allows us to
give a reasonable no-tails characterisation that may hold even in cases
in which HP does not. Actually, we shall see in section 5 that CPP, although
in principle weaker than HP and TF, in practice always implies them, except
in two spacetime dimensions. We wish to stress that the restriction of
data we are considering is actually a very small one, concerning a subset 
of $\cal C$ of measure zero which contains points that are already
pathological. Furthermore, such a restriction corresponds precisely to
what is commonly done when assigning data on the asymptotic past (probably
the only characteristic initial value problem with good physical
motivations), where one allows $\phi\ne0$ at past null infinity ${\cal
I}^-$ but requires $\phi=0$ at past timelike infinity $i^-$ [14].

In addition, the notion of progressing waves motivates
the following definition for two-dimensional equations:

\vskip2pt
\leftskip=20pt
\noindent{\bf PW$\!_N$:} A wave equation in two spacetime dimensions is
said to satisfy the {\it progressing wave propagation property of order\/}
$N$ (PW$\!_N$) iff its general solution can be written as a sum of two
progressing waves,
  $$
  \phi(x) = \sum_{i=0}^N f_i(x)\,R^{(i)}(u(x))
  + \sum_{i=0}^N g_i(x)\,S^{(i)}(v(x))\;, \eqno(3.3)
  $$
where the amplitudes $f_i$ and $g_i$ are fixed functions on $\cal M$
depending on the wave equation (and at least one of $f_N$ and $g_N$ is
not identically zero), while the wave forms $R$ and $S$ are arbitrary
sufficiently differentiable functions of one variable (this forces the
coordinates $u$ and $v$ to be null) [17,19,22].

\leftskip=0pt
\vskip3pt
\noindent In section 6 we shall show that, in two spacetime dimensions,
CPP is equivalent to PW$\!_0$; as far as the PW$\!_N$, with $N>0$, are
concerned, their resemblance to a no-tails property is only apparent. 
Furthermore, wave equations in more than two dimensions may well have
progressing wave solutions for appropriate choices of the wave front (see,
e.g., Ref 16), but it seems excessive to ask that their {\it general\/}
solution be expressible as a finite sum of progressing waves.  These issues,
which diminish the appeal of PW$\!_N$ as regards the study of tails, will be
discussed in the concluding section.

Another property that can be satisfied by two-dimensional wave equations
is their solvability by the method of Kundt and Newman [20]. To apply this
method, we start by writing (2.7) in the two equivalent normal forms
  $$\eqalignno{
  &\de_v(j_0\,\de_u\,\phi_0)-j_1\,\phi_0=0\;, &(3.4)\cr
\noalign{\smallskip}
  &\de_u(l_0\,\de_v\,\psi_0)-l_{-1}\,\psi_0=0\;, &(3.5)}
  $$
where $\phi_0$ and $\psi_0$ are related to $\phi$ by factor
transformations $\phi_0=\phi\exp\sigma$ and $\psi_0=\phi\exp\tau$,
with
  $$
  \sigma(u,v):=\int^u\dd u'\,V(u',v)\,,
  \eqno(3.6)
  $$
  $$
  \tau(u,v):=\int^v\dd v'\,U(u,v')\,,
  \eqno(3.7)
  $$
while the coefficients are related to those in (2.7) by
  $$
  j_0=l_0^{-1}=\exp(\tau-\sigma)\,,
  \eqno(3.8)
  $$
  $$
  j_1=(\de_vV+UV-W)\,j_0\,,
  \eqno(3.9)
  $$
and 
  $$
  l_{-1}=(\de_uU+UV-W)\,l_0\;.
  \eqno(3.10)
  $$
If we inductively define
$j_k$ and $\phi_k$ by
  $$\eqalignno{
  &j_{k+1}/j_k = j_k/j_{k-1} - \de^2_{uv} \ln|j_k|\;, &(3.11)\cr
\noalign{\smallskip}
  &j_{k+1}\phi_{k+1} = j_k\,\de_u\phi_k\;, &(3.12)}   
  $$
assuming of course $j_k\not=0$ for all $k\in\ZZ$, we obtain from
(3.4) a countable set of equations 
  $$
  \de_v(j_k\de_u\,\phi_k)-j_{k+1}\,\phi_k = 0\;, \qquad k\in\ZZ\;.
  \eqno(3.13)
  $$
Similarly, if we inductively define $l_k$ and $\psi_k$ by
  $$\eqalignno{
  &l_{k-1}/l_k = l_k/l_{k+1} - \de^2_{vu} \ln|l_k|\;, &(3.14)\cr
\noalign{\smallskip}
  &l_{k-1}\psi_{k-1} = l_k\,\de_v\psi_k\;, &(3.15)}
  $$
assuming now that $l_k\not=0$ for all $k\in\ZZ$, we obtain from
(3.5) a second countable set
  $$
  \de_u(l_k\de_v\,\psi_k)-l_{k-1}\,\psi_k = 0\;, \qquad k\in\ZZ\;.
  \eqno(3.16)
  $$
We shall refer to equations (3.13) and (3.16) as being in the $k^{\rm
th}$ $v$- and $u$-normal form, respectively. It is not hard to check that
for all $k\in\ZZ$ the $k^{\rm th}$ $v$-normal form equation, corresponding
to the coefficients $j_k$ and $j_{k+1}$, and the $k^{\rm th}$ $u$-normal
form equation, corresponding to the coefficients $l_k$ and $l_{k-1}$, are
equivalent under the transformation
  $$\eqalignno{
  &j_k\,l_k = 1\;,   &(3.17)\cr
\noalign{\smallskip}
  &\phi_k = l_k\,\psi_k\;, &(3.18)}
  $$ 
for $k\in\ZZ$. It is also easy to see that the equations within the set
(3.13) (respectively, (3.16)) are equivalent in the sense that a
solution of any one of them generates a solution of every other one of them
through (3.11)--(3.13), (3.17), and (3.18) (respectively,
(3.14)--(3.18)), and we thus obtain two equivalence classes of $v$- and
$u$-normal form equations labelled by the index $k$ ranging over both
negative and positive integers.

Given a wave equation in the $0^{\rm th}$ $v$-normal form (3.4), we
say that its substitution sequence $\{j_k\}$ is {\it double terminating in
$N$ steps\/} [20] when $j_{k_1+1}=0$ and $l_{k_2-1}=0$ (but $j_{k_1}\not=0$
and $l_{k_2}\not=0$) for some $k_1\ge0$, $k_2\le0$, and
$N=\max\,\{k_1,-k_2\}$. However, it was shown in Ref 20, using
(3.11)--(3.18), that in this case the general solution of (3.4) is
$\phi_0=\phi_A+l_0\phi_R$, where
  $$\eqalignno{
  \phi_A&:= {1\over j_1}\,\de_v\biggl( {j_1\over j_2}\,\de_v \biggl(
  {j_2\over j_3}\,\de_v\biggl( \cdots {j_{k_1-1}\over j_{k_1}}\,\de_v
  \biggl(j_{k_1}\,S(v)\biggr)\cdots\biggr)\biggr)\biggr)\;,
  &(3.19)\cr
  \phi_R&:= {1\over l_{-1}}\,\de_u \biggl( {l_{-1}\over l_{-2}}\,\de_u 
  \biggl({l_{-2}\over l_{-3}}\,\de_u\biggl(\cdots{l_{k_2+1}\over l_{k_2}}\,
  \de_u\biggl(l_{k_2}\,R(u)\biggr)\cdots\biggr)\biggr)\biggr)\;, 
  &(3.20)}
  $$
with $S(v)$ and $R(u)$ arbitrary functions of one variable. It is obvious
from (3.19) and (3.20) that such a $\phi$ is a progressing wave of
order $N$. This relationship motivates the following definition:
\vskip2pt
\leftskip=20pt
\noindent{\bf KN$\!_N$:} A wave equation in two spacetime dimensions is
said to be {\it solvable by the Kundt-Newman method in $N$ steps\/}
(KN$\!_N$) iff its substitution sequence is double terminating in $N$
steps.

\leftskip=0pt
\vskip3pt
\noindent As we have just seen, all KN$\!_N$ wave equations are
PW$\!_N$. We shall see in section 7 that the converse is also true, so
that KN$\!_N$ and PW$\!_N$ are equivalent properties. 


\goodbreak
\section{4.}{Equivalence between HP and TF}

\noindent We now begin studying the relationships between the various
properties of wave equations we listed in the previous section. The first
three are related in a simple way in any number of dimensions; as we will
now show, TF and HP are always trivially equivalent, which justifies the
fact that they are often identified, and they are almost always equivalent
to CPP (see next section). These results generalise to arbitrary wave
equations of the type (1.3) those of Ref 15.

In order to show that TF $\Rightarrow$ HP, let us consider any point $x\in
D^+(\cal S)$, and assign data with support ${\cal C}\subset I^-(x)\cap\cal
S$. Since TF holds by hypothesis, and $x\in K^+(\cal C)$, it follows that
$\phi(x)=0$ for each set of data prescribed on such a $\cal C$. By
causality, $\phi(x)$ cannot be influenced by data given outside
$I^-(x)\cap\cal S$, hence it can depend only on their value on
$E^-(x)\cap\cal S$, where $E^-(x)$ stands for the set of points in
$J^-(x)$ that are null related to $x$. This is precisely the content of HP.

Let us now prove that HP $\Rightarrow$ TF. First of all, let us notice
that, by choosing data that vanish everywhere on $\cal S$, we get 
$\phi(x)=0$ everywhere in $\cal M$ by (1.3). If we now assign
data on $\cal S$ with compact support $\cal C$, HP implies that the value of
$\phi(x)$, for any $x\in K^+(\cal C)$, must be independent of the data;
in particular, it must have the same value than in the case of vanishing
data, i.e., it must be equal to zero, by the remark above. This
completes the proof that HP $\Leftrightarrow$ TF.

It is interesting to notice that the structure of this proof allows one to
extend it to more general wave equations than (1.3). Actually, in the
case of the latter, a simpler proof can be given that makes use of the
equivalence between HP and the property $\tw\Delta(x',x)\equiv 0$. In
fact, it follows immediately from (3.1) that, if HP is satisfied, then
$\phi(x)=0$ for all $x\in K^+({\cal C})$, i.e., the wave equation is TF. 
The converse is also true, as we can see by choosing $\cal S$ to be
spacelike at a point $\bar x$, and data of support only at $\bar x$, such
that, denoting by $n^a$ the unit vector normal to
$\cal S$,
  $$\eqalignno{
  &\phi|_{\cal S}(x) \equiv 0\;, &(4.1)\cr
  \noalign{\smallskip}
  &(n^a\nab_a\phi)|_{\cal S}(x) = \delta_{\cal S}(x,\bar
  x)\;, &(4.2)}
  $$
where $\delta_{\cal S}$ is the $(m-1)$-dimensional delta function on $\cal
S$. Then (3.1) reduces to
  $$
  \phi(x) = -\Delta(\bar x,x)\;, \eqno(4.3)
  $$
and if $\phi(x)=0$ for all $x\in K^+({\cal C})=I^+(\bar x)$, we must have
$\Delta(\bar x,x)=0$ for all such pairs of points, i.e., 
$\tw\Delta(x',x)\equiv0$. However, this proof relies heavily on the
existence of the integral representation (1.8), and appears much more
difficult to generalise to the case of more complicated wave equations
than the ``geometrical'' one given above.


\goodbreak
\section{5.}{Relationship between HP and CPP}

\noindent It is obvious from the definitions given in section 3 that TF, and
thus HP, imply CPP. To check whether the converse holds we need to 
consider again the integral expression (3.1) for $x\in K^+({\cal C})$,
where $\cal C$ is now part of a piecewise null hypersurface $\cal S$ locally
defined by the condition $w=\rm const$, for some $w:{\cal M}\rightarrow{\Bbb
R}$ with $g^{ab}\,\nab_aw\,\nab_bw=0$ at all points where $\cal S$ is
differentiable (this condition may fail only in a subset of $\cal S$
of measure zero). One usually writes $\dd S_a=\dd S\,n_a$, where $\dd S$
and $n_a$ are often thought of as volume element and unit normal to the
surface $\cal S$, respectively, which clearly does not make sense if $\cal
S$ is null. However, the definitions of these objects can be extended to
that case as well. Let us introduce local coordinates $\{\xi^i:
i=1,\ldots,m-1\}$ on $\cal S$, and use $w$ as a coordinate transverse to
$\cal S$, so that $(w,\xi^i)$ are coordinates on $\cal M$ adapted to $\cal
S$. Then we can always write
$\dd S_\mu = \dd^{m-1}\xi\,\sqrt{-g}\,{\delta_\mu}^w$, with the understanding
that $g$ must be calculated in these coordinates, which in the null case
can be split into $\dd S=\dd^{m-1}\xi\,\sqrt{-g}$, and
$n_\mu={\delta_\mu}^w$; the latter are the components of the form
$n_a=\nab_aw$. With these conventions, we can write the Gauss theorem for
an arbitrary vector field $Y^a$ as
  $$
  \int_{\cal N}\dd V\,\nab_aY^a=\oint_{\de\cal N}\dd S_a\,Y^a\;,
  \eqno(5.1)
  $$
where $\cal N$ is a region in $\cal M$, and $\dd V=\dd^mx\sqrt{-g}$ is
the spacetime volume element, even when (part) of $\de\cal N$ is
null. This follows from the fact that the left hand side of (5.1) can
be broken into a sum of integrals of the form
  $$
  \int_{\cal D}\dd^{m-1}\xi\,\dd w\,\de_\mu\left(\sqrt{-g}\,Y^\mu\right)\;,
  \eqno(5.2)
  $$
where $\cal D$ is a domain in $\IR^m$ one of whose ``faces'' (that we
denote by $\cal F$) is defined by $w=0$; but (5.2) can be
transformed, using the Gauss theorem in $\IR^m$, into a sum of terms
corresponding to the various faces of $\cal D$, of which only 
  $$
  \int_{\cal F}\dd^{m-1}\xi\,\sqrt{-g}\,{\delta_\mu}^w\,Y^\mu 
  \eqno(5.3)
  $$
contributes, in the end, to the total expression. Therefore, (5.1)
(hence (1.8) and (3.1)) is completely justified, independently of the
causal character of $\de\cal N$.

We can rewrite (3.1) in the coordinates $(w,\xi^i)$ as
  $$
  \phi(x) = -\int_{\cal C}\dd S' \Bigl\{\nabp_\mu
  (n^\mu\Delta\phi) - \phi\,\bigl[(\nabp_\mu n^\mu)\Delta
  + n^\mu (2\,\nabp_\mu\Delta - H_\mu\Delta)\bigr]
  \Bigr\}\,\theta(-\Gamma)
  \eqno(5.4)
  $$
(here and in the following integrals, all functions in the integrand depend
on $x'$, and all two-point functions on the ordered pair $(x',x)$). Since
we have $n^\mu=g^{\mu w}$, which implies $n^w=g^{ww}=g^{\mu\nu}\de_\mu
w\,\de_\nu w=0$, the first term in the right hand side of (5.4) contains
only derivatives performed along
$\cal S$, and can be transformed as
  $$
  -\int_{\cal C}\dd S'\,\nabp_\mu\left(n^\mu\,\Delta\,\phi\right)=
  -\int_{\cal F}\dd^{m-1}\xi'\,\de'_i\left(\sqrt{-g}
  \,n^i\,\Delta\,\phi\right)=
  -\oint_{\de\cal C}\dd\sigma'_a\,n^a\,\Delta\,\phi\,,
  \eqno(5.5)
  $$
where $\cal F$ stands again for the region of $\IR^m$ defined
by $w=0$, $\dd\sigma_a$ is the oriented volume element on
$\de\cal C$, and the Gauss theorem in $\IR^{m-1}$ has been applied in the
last step (possible discontinuities of $\phi$ do not prevent us from using
the theorem, if they are correctly interpreted in the sense of
distributions). We can therefore deduce that this term does not contribute
to the expression, because $\phi(x')=0$ outside $\cal C$ and we can think
of the integration as being performed over a region larger than $\cal C$,
on whose boundary $\phi(x')=0$. Therefore, CPP is satisfied iff
  $$
  \int_{\cal C}\dd S'\,\phi\bigl[ (\nabp_\mu n^\mu)\,\Delta
  + n^\mu\,(2\,\nabp_\mu\Delta-H_\mu\Delta) \bigr]\,\theta(-\Gamma) = 0
  \eqno(5.6)
  $$
for all data on an arbitrary compact null $\cal C$. In particular,
we can choose as data
  $$
  \phi|_{\cal S}(x) = \delta_{\cal S}(x,\bar x)\;, \eqno(5.7)
  $$
for any nonsingular point $\bar{x}\in\cal C$. Since $\bar x$ is arbitrary,
this implies that either $\Delta(x',x)=0$, and we get HP directly, or
  $$
  \nab_an^a(x)+n^a(x)
  \bigl[2\,\nab_a\ln|\Delta(x,x')|-H_a(x)\bigr]=0 
  \eqno(5.8)
  $$
for all pairs of points with $x'\in I^+(x)$, $x\in\cal C$ nonsingular, and
all null vector fields $n^a$ which are gradients of a function.

We will now study the consequences of (5.8), by writing it in
a suitable coordinate system. In terms of $w$, (5.8) is of the form
  $$
  g^{ab}\nab_a\nab_b w + X^a\nab_a w = 0\;, \eqno(5.9)
  $$
where $X^a(x,x') := 2\,\nabla^a\ln|\Delta(x,x')|-H^a(x)$. Choose any
point $x_0\in\cal C$, and a system of Riemann normal coordinates based at
$x_0$, for which $g_{\mu\nu}(x) = \eta_{\mu\nu} + {\cal O}(2)$, where
we have introduced the  notation ${\cal O}(n)$ for a quantity which is of
order $n$ in the coordinate separation of $x$ from $x_0$; then (5.9)
becomes
  $$
  \eta^{\mu\nu}\de_\mu\de_\nu w + X^\mu\de_\mu w = {\cal O}(1)\;.
  \eqno(5.10)
  $$
Let us now consider the function
  $$
  w = t + {\textstyle\sqrt{x^ix_i}} + c_{\mu\nu}x^\mu x^\nu+{\cal O}(3)\;,
  \eqno(5.11)
  $$
where the $c_{\mu\nu}$ are appropriate coefficients such that
$g^{\mu\nu}\de_\mu w\,\de_\nu w=0$. Substituting into
(5.10) we get that, for $x^ix_i\neq0$, 
  $$
  {m-2\over\sqrt{x^ix_i}}+2\,\eta^{\mu\nu}c_{\mu\nu}+
  X^t(x,x')+{\eta_{ij}X^i(x,x')x^j\over\sqrt{x_k x^k}}={\cal O}(1)\;.
  \eqno(5.12)
  $$
The first term in the left hand side of (5.12) is ${\cal O}(-1)$, whereas
the others are all ${\cal O}(0)$; therefore (5.12) implies $m=2$ and
  $$
  X^t(x_0,x')+X^x(x_0,x')\lim_{x\rightarrow 0^\pm}{x\over |x|}=
  -2\,\eta^{\mu\nu}c_{\mu\nu}\;,
  \eqno(5.13)
  $$
where the $\pm$ in the limit corresponds to the arbitrariness in the
direction of approach of $x^\mu$ to the origin and, with some abuse of
notation, we have denoted by $x$ the spatial coordinate of the
two-dimensional Minkowski spacetime. This means that
  $$
  X^t(x_0,x')\pm X^x(x_0,x')=-2\,\eta^{\mu\nu}c_{\mu\nu}
  \eqno(5.14)
  $$
must hold simultaneously for both signs, i.e., that $X^x(x_0,x')=0$. 
Since this is true in any system of local Minkowskian coordinates, and
taking into account the arbitrariness of $x_0$, we deduce that 
$X^a(x,x')=0$ as a field.

In conclusion, we have found that CPP implies either HP, or $m=2$ and 
  $$
  H_a(x) = 2\,\nab_a\ln |\Delta(x,x')|\;.
  \eqno(5.15)
  $$
In other words, CPP is always equivalent, for $m>2$, to HP; in two
spacetime dimensions $\Delta\neq0$ and HP is always violated, whereas CPP
can be satisfied. Therefore, CPP can be regarded as a nontrivial
generalisation of HP, although it becomes interesting by itself only for
$m=2$. More explicitly, the definitions of CPP and HP differed in that for
the former, the support of the data had to be null rather than achronal, and
the data had to vanish at non-smooth points of $\cal C$. We now know that
this is of no consequence other than in some two-dimensional cases, but it is
perhaps surprising that such a small restriction in the data can have the
effect of enlarging, in a nontrivial way, the class of equations to which
the definition applies.

We have derived (5.15) above as a necessary condition for the validity of
CPP in two dimensions, but the following simple argument shows that it is
also sufficient. Substituting (5.15) into (5.9) the latter reduces to
$g^{ab}\nab_a\nab_bw=0$, and CPP holds if this equation is satisfied by any
function $w$ that generates null hypersurfaces. For $m=2$ this is true,
because $w$ is null iff it depends on only one of the coordinates $u$ and
$v$ defined by (2.5) and (2.6), i.e., iff either $w=w_1(u)$ or $w=w_2(v)$. 
Using the identity (2.2) for $f=w$, we have $g^{ab}\nab_a\nab_bw=0$. 

Let us now see what restrictions the condition (5.15) places on the wave
equation. First of all, since the Riemannian connection is torsion-free,
we have $\nab_{[a}\nab_{b]}\ln |\Delta|=0$, and (5.15) implies
  $$
  \nab_{[a}H_{b]}=0\;;
  \eqno(5.16)
  $$
this is an integrability condition that allows us to obtain $\Delta$ by
straightforward integration of (5.15). In fact, (5.16) implies that $H_a =
2\,\nab_a\Lambda$, for some $\Lambda$; substituting back into (5.15), we have
  $$
  |\Delta(x,x')|={\exp\Lambda(x)\over\exp\Lambda(x')}\;,
  \eqno(5.17)
  $$
since $\Delta(x',x')=1$ by (2.9) and (2.10). Furthermore, it is
not difficult to verify, from the definition of the Green function, that
$\Delta(x,x')$ satisfies the ``adjoint wave equation'' in $x$ (see also
(2.8) and (2.11))
  $$
  g^{ab}\nab_a\nab_b\Delta-\nab_a\left(H^a\Delta\right)+K\,\Delta=0\;.
  \eqno(5.18)
  $$
Substituting (5.15) twice into (5.18), we get
  $$
  2\,\nab_aH^a+g_{ab}H^aH^b-4\,K=0\;,
  \eqno(5.19)
  $$
hence
  $$
  g^{ab}\nab_a\nab_b\Lambda+g^{ab}\nab_a\Lambda\nab_b\Lambda-K=0\;.
  \eqno(5.20)
  $$
Using (5.17), we can replace (5.20) by an equation for $\Delta$ simpler than
(5.18),
  $$
  g^{ab}\nab_a\nab_b\Delta-K\Delta=0\;,
  \eqno(5.21)
  $$
which can also be directly obtained by using (5.19) into (5.18). Equations
(5.16) and (5.20) (or (5.19)) are also sufficient conditions for CPP, as one
can easily see by considering a new field $\psi:=\phi\exp\Lambda$. 
From (1.3) and (5.20), we find that $\psi$ obeys the equation
$g^{ab}\nab_a\nab_b\psi=0$, which is CPP because 
$m=2$. Since $\psi$ and $\phi$ differ only by the nonvanishing
factor $\exp\Lambda$, it follows that $\phi$ satisfies CPP as well.

In the coordinates $(u,v)$ we have $H^u=-\Omega^{-2}U$, $H^v=-\Omega^{-2}V$,
$K=-\Omega^{-2}W$, so that (5.16) and (5.19) are equivalent to 
  $$
  \de_uU=\de_vV \eqno(5.22)
  $$ 
and
  $$
  \de_uU +  UV - W = 0\;, \eqno(5.23)
  $$ 
respectively. Similarly, we have $U=\de_v\Lambda$ and $V=\de_u\Lambda$;
(5.20) and (5.21) become
  $$
  \de^2_{uv}\Lambda+\de_u\Lambda\,\de_v\Lambda-W=0
  \eqno(5.24)
  $$
and
  $$
  \de^2_{uv}\Delta-W\Delta=0\;.
  \eqno(5.25)
  $$
These conditions for CPP represent strong constraints on the wave equation;
for example, when $U=V=0$, (5.23) and (5.24) are satisfied only if $W=0$
[15]. Notice that when (5.16), or equivalently (5.22), is satisfied, the
functions $\sigma$ and $\tau$ defined in (3.6) and (3.7) both reduce to
$\Lambda$; upon substitution into (3.8)--(3.10), equations (5.22) and (5.23)
then imply that $j_0=l_0=1$ and $j_1=l_{-1}=0$.


\goodbreak
\section{6.}{Equivalence between CPP and PW$\!_0$ in two spacetime 
dimensions}

\noindent In this section we will show that, in two spacetime dimensions,
CPP and PW$\!_0$ are equivalent properties; the proof is very simple if
we rely on some results of the previous section. In fact, since the
two-dimensional wave equation (2.7) satisfies CPP iff $\phi$ can be
written as $\phi=\exp(-\Lambda)\,\psi$, with $\de^2_{uv}\psi=0$, it follows
that (2.7) is CPP iff its general solution has the form
  $$
  \phi(u,v)=\ee^{-\Lambda(u,v)}\bigl(r(u)+s(v)\bigr)\;,
  \eqno(6.1)
  $$ 
where $r$ and $s$ are arbitrary functions. By comparing (6.1) with
the $N=0$ case of (3.3), 
  $$
  \phi(u,v)=f_0(u,v)\,R(u)+g_0(u,v)\,S(v)\;,
  \eqno(6.2)
  $$ 
we conclude immediately that CPP $\Rightarrow$ PW$\!_0$. In order to prove
the converse, i.e., that PW$\!_0\Rightarrow$ CPP, substitute the expression
(6.2) into (2.7); invoking the arbitrariness of $R$ and $S$, we
get:
  $$\eqalignno{
  \de^2_{uv}f_0+U\de_uf_0&+V\de_vf_0+Wf_0=0\;; &(6.3) \cr
\noalign{\smallskip}
  \de^2_{uv}g_0+U\de_ug_0&+V\de_vg_0+Wg_0=0\;; &(6.4) \cr
\noalign{\smallskip}
  \de_vf_0&+Uf_0=0\;; &(6.5) \cr
\noalign{\smallskip}
  \de_ug_0&+Vg_0=0\;. &(6.6) }
  $$ 
After some manipulations, these equations lead to 
(5.22) and (5.23), which are sufficient conditions for the validity of
CPP, as we saw in section 5. We can therefore conclude that, in a
two-dimensional spacetime, CPP and PW$\!_0$ are equivalent.

An immediate corollary of this result is that not every expression of the
type (6.2) is the general solution of a two-dimensional wave equation. 
This is evident by a comparison between (6.1) and (6.2), which shows that
the amplitudes $f_0$ and $g_0$ must be related by
$f_0(u,v)/g_0(u,v)=\alpha(u)/\beta(v)$, for some functions $\alpha$ and
$\beta$. The same conclusion can be reached in a more explicit way by
integration of (6.5) and (6.6), to obtain
  $$\eqalignno{
  f_0(u,v) &= f_0(u,0)\,\exp\left(-\int^v_0\dd v'\,U(u,v')\right)=
  \alpha(u)\,\exp\bigl(-\Lambda(u,v)\bigr)\;,
  &(6.7) \cr
  g_0(u,v) &= g_0(0,v)\,\exp\left(-\int^u_0\dd u'\,V(u',v)\right)=
  \beta(v)\,\exp\bigl(-\Lambda(u,v)\bigr)\;,
  &(6.8) }
  $$ 
with $\alpha(u):=f_0(u,0)\exp\Lambda(u,0)$ and $\beta(v):=
g_0(0,v)\exp\Lambda(0,v)$. By a generalisation of this argument, one could
show that not every expression of the type (3.3), with arbitrary $R$ and $S$,
is the general solution of a two-dimensional wave equation.

It is possible to prove that CPP $\Leftrightarrow$ PW$\!_0$ without
necessarily using the results of the previous section. This is obvious for
what concerns the implication PW$\!_0$ $\Rightarrow$ CPP, since PW$\!_0$ 
leads us to the same equations (5.22) and (5.23) of section 5, and one can
proceed to show that they are sufficient conditions for CPP exactly in the
same way as we did there. The proof that CPP $\Rightarrow$ PW$\!_0$ is less
compact, but we nevertheless present it because of its intrinsic interest.

Consider the null data of compact support given by $\varphi(u) =
\delta(u-u_0)$, with $u_0>0$, and $\psi(v) \equiv 0$; then CPP implies that
$\phi(u,v)$ must be concentrated on the line $u=u_0$. This means that, for
any fixed $v$, $\phi(u,v)$ must be a distribution in $u$ that, applied to a
test function $F(u,v)$, produces a number depending only on the behaviour of
$F$ at the point $(u_0,v)$, i.e., on $F^{(i)}(u_0,v)$, with $i\geq0$. Since
distributions are linear functionals, the combination of the various
$F^{(i)}(u_0,v)$ must be linear, so that we can write
  $$
  \phi(u,v) =
  \sum_{i=0}^{+\infty}f_i(u_0,v)\,\delta^{(i)}(u-u_0)\;, \eqno(6.9)
  $$
for some functions $f_i$. The sum on the right hand side of (6.9) actually
consists only of a finite number of terms; this follows from a rigorous
result of distribution theory [23], but can be heuristically justified as
follows. Since the action of $\phi$ as a functional must be defined on any
test function $F$, the value of the series
  $$
  \sum_{i=0}^{+\infty}f_i(u_0,v)F^{(i)}(u_0,v) \eqno(6.10)
  $$
must be finite for each $F$. Suppose that the sum in (6.9) and (6.10) is
infinite, that is, that there exist infinitely many  $f_i\neq0$. Then,
taking $F$ such that $F^{(i)}(u_0,v)=f_i(u_0,v)^{-1}$ when $f_i\neq0$,
and arbitrary otherwise, the expression (6.10) becomes infinite, which
contradicts the hypothesis of regularity of $\phi$. Hence, we must have
  $$
  \phi(u,v)=\sum_{i=0}^{N_u}f_i(u_0,v)\,\delta^{(i)}(u-u_0)\;,
  \eqno(6.11)
  $$
with $N_u$ finite. Similarly, if $\varphi(u)\equiv 0$ and
$\psi(v) = \delta(v-v_0)$, we have
  $$
  \phi(u,v)=\sum_{i=0}^{N_v}g_i(u,v_0)\,\delta^{(i)}(v-v_0)\;,
  \eqno(6.12)
  $$
for some finite $N_v$ and some functions $g_i$.

From (6.11) and (6.12) it already follows that (2.7)
is 
PW$\!_N$, with $N=\max\,\{N_u,N_v\}$, as can be seen just by using
linearity and representing arbitrary data $\varphi$ and $\psi$ as
superpositions of $\delta$-like data,
  $$\eqalignno{
  &\varphi(u) = \int\dd u_0\,\varphi(u_0)\,\delta(u-u_0)\;, &(6.13)\cr
  &\psi(v) = \int\dd v_0\,\psi(v_0)\,\delta(v-v_0)\;, &(6.14)}
  $$
but without making any other use of the differential equation for $\phi$.
We shall now show, however, that (2.7) allows us to set $N=0$, i.e., to
conclude that CPP $\Rightarrow$ PW$\!_0$. For this purpose, let us first
consider again data $\varphi(u)=\,\delta(u-u_0)$, $\psi(v)\equiv 0$.
Substituting the solution (6.11) into (2.7) we get
  $$\eqalignno{
  &\Bigl[\de_vf_{N_u}(u_0,v)+U(u,v)f_{N_u}(u_0,v)\Bigr]\,
  \delta^{(N_u+1)}(u-u_0)\ +\cr
  &+\sum_{i=1}^{N_u}\Bigl[V(u,v)\,\de_vf_i(u_0,v)+W(u,v)f_i(u_0,v)\ + \cr
  &+\de_vf_{i-1}(u_0,v)+U(u,v)f_{i-1}(u_0,v)\Bigr]\,\delta^{(i)}(u-u_0)\
  +\cr &+\Bigl[V(u_0,v)\,\de_vf_0(u_0,v)+W(u_0,v)\,f_0(u_0,v)\Bigr]\,
  \delta(u-u_0)=0\;.
  &(6.15)}
  $$
Smoothing (6.15) with a test function, and taking into account the
arbitrariness of the latter, we obtain the following set of equations:
  $$\eqalignno{
  &\de_vf_{N_u}(u_0,v)+U(u_0,v)\,f_{N_u}(u_0,v)=0\;; &(6.16) \cr
  \noalign{\medskip}
  &\sum_{i=k}^{N_u}(-1)^i{i\choose k}\Bigl[\de_u^{i-k}V(u_0,v)\,
  \de_vf_i(u_0,v)+\de_u^{i-k}W(u_0,v)\,f_i(u_0,v)\ + \cr
  &\quad+\de_u^{i-k}U(u_0,v)\,f_{i-1}(u_0,v)\Bigr]
  +(-1)^k\de_vf_{k-1}(u_0,v)\ + \cr
  &\quad +\ (-1)^{N_u+1}{N_u+1\choose k}\de_u^{N_u-k+1}U(u_0,v)\,
  f_{N_u}(u_0,v) = 0\;; \qquad N_u\geq k\geq 1\;; &(6.17) \cr
  \noalign{\medskip}
  & \sum_{i=1}^{N_u}(-1)^i\Bigl[\de_u^iV(u_0,v)\,\de_vf_i(u_0,v) +
  \de_u^iW(u_0,v)\,f_i(u_0,v)+ \de_u^iU(u_0,v)\,f_{i-1}(u_0,v)\Bigr]\ + \cr
  &\quad +\ V(u_0,v)\,\de_vf_0(u_0,v)+W(u_0,v)\,f_0(u_0,v)=0\ + \cr
  &\quad +\ (-1)^{N_u+1}\,\de_u^{N_u+1}U(u_0,v)\,f_{N_u}(u_0,v)\;.
  \vphantom{\sum_{i=1}^n}&(6.18) }
  $$
From (6.16) and the fact that $f_{N_u}(u_0,0)=0$, which follows from using
our data in (6.11), we obtain that $f_{N_u}(u_0,v)=0$ for all $v$. Using
this in (6.17) and repeating the procedure, we obtain $f_i(u_0,v)=0$,
$\forall i\geq1$, so that only $f_0(u_0,v)$ can be nonzero. An analogous
argument, using data $\varphi(u)\equiv0$, $\psi(v)=\delta(v-v_0)$, shows
that the only nonvanishing coefficient in (6.12) can be $g_0(u,v_0)$. 
Therefore, the general solution, obtained from (6.11) and (6.12) using
arbitrary initial data as in (6.13) and (6.14), is
  $$
  \phi(u,v) = f_0(u,v)\,\varphi(u) + g_0(u,v)\,\psi(v)\;, \eqno(6.19)
  $$
from which the validity of PW$\!_0$ follows. Furthermore, it is not
difficult to see that (6.17) for $k=1$ reduces to (6.5) in $u=u_0$,
and that substituting this relation into (6.18) we obtain (5.23). By a
completely symmetric procedure we can recover (6.6).


\goodbreak
\section{7.}{Equivalence between PW$\!_N$ and KN$\!_N$}

\noindent As we saw in section 3, all KN$\!_N$ wave equations are PW$\!_N$;
it has been conjectured that the converse is also true, namely that all
PW$\!_N$ equations can in fact be solved exactly by the Kundt-Newman method
in $N$ steps [20,24]. To show that this is indeed the case, suppose we have a
2-dimensional wave equation whose general solution is the  progressing wave
(3.3). By suitable transformations, we can always write the wave equation in
its $0^{\rm th}$ $v$- or $u$-normal form, (3.4) or (3.5); let us concentrate
on the $v$-normal form.

Applying the differential operator in (3.4) to the progressing wave
(3.3), and imposing that the coefficients of derivatives of the arbitrary
functions $R(u)$ and $S(v)$ of different orders vanish separately in
the resulting expression, we have:
  $$\eqalignno{
  &\de_v(j_0\,\de_u\,f_0)-j_1\,f_0=0\;, &(7.1) \cr
\noalign{\smallskip}
  &\de_v(j_0\,\de_u\,f_i)-j_1\,f_i+\de_v(j_0\,f_{i-1})=0\;,
  \qquad 1\le i\le N\;, &(7.2) \cr
\noalign{\smallskip}
  &\de_v(j_0\,f_N)=0\;, &(7.3) \cr
  }$$
and
  $$\eqalignno{
  &\de_v(j_0\,\de_u\,g_0)-j_1\,g_0=0\;. &(7.4) \cr
\noalign{\smallskip}
  &\de_v(j_0\,\de_u\,g_i)-j_1\,g_i+j_0\,\de_ug_{i-1}=0\;,
  \qquad 1\le i\le N\;, &(7.5) \cr
\noalign{\smallskip}
  &\de_ug_N=0\;, &(7.6) \cr
  }$$

Let us consider the group (7.4)--(7.6). Equations (7.4) and (7.5) can be
rewritten more conveniently as 
  $$
  \de_v\biggl({j_0\over j_1}\,\de_ug_0\biggr)+\de_v\ln |j_1|\,{j_0\over
  j_1}\,\de_ug_0-g_0=0
  \eqno(7.7)
  $$
and
  $$
  \de_v\biggl({j_0\over j_1}\,\de_ug_i\biggr)+\de_v\ln |j_1|\,{j_0\over
  j_1}\,\de_ug_i-g_i+{j_0\over j_1}\,\de_ug_{i-1}=0\,, \qquad 1\le i\le N\,,
  \eqno(7.8)
  $$ 
respectively. By repeated differentiation of (7.8) with respect to $u$,
one finds that, for any $h\geq 1$ and $k=0,\ldots,N-1$,
  $$
  \de_v{\cal D}_u^h\,g_{N-k}+\de_v\ln |j_h|\,{\cal D}_u^h\,g_{N-k}-
  {\cal D}_u^{h-1}g_{N-k}+{\cal D}_u^h\,g_{N-k-1}=0\;,
  \eqno(7.9)
  $$
where the differential operator ${\cal D}_u^k$, containing $k$ derivatives
$\de_u$, is defined as
  $$
  {\cal D}_u^k\,F:={j_{k-1}\over j_k}\,\de_u\biggl({j_{k-2}\over
  j_{k-1}}\,\de_u \biggl(\cdots{j_1\over j_2}\,\de_u\biggl({j_0\over
  j_1}\,\de_uF\biggr)\cdots\biggr)\biggr)\;, \eqno(7.10)
  $$  
for any sufficiently differentiable function $F$. We shall prove (7.9) by
induction over $h$. Assuming that it holds for some $h$, and
differentiating it with respect to $u$, we have
  $$
  \de^2_{uv}{\cal D}_u^h\,g_{N-k}+\de_v\ln
  |j_h|\,\de_u{\cal D}_u^h\,g_{N-k}-{j_{h+1}\over j_h}\,{\cal
  D}_u^{h}g_{N-k}+\de_u{\cal D}_u^{h}g_{N-k-1}=0\;,
  \eqno(7.11)
  $$
where we have used (3.11). Rewriting the first term as
  $$
  \de_v\biggl({j_{h+1}\over j_h}\,{j_h\over j_{h+1}}\,
  \de_u{\cal D}_u^h\,g_{N-k}\biggr) = \de_v\ln
  \bigg|{j_{h+1}\over j_h}\biggr|\,\de_u{\cal
  D}_u^h\,g_{N-k}+{j_{h+1}\over j_h}\,\de_v{\cal
  D}_u^{h+1}g_{N-k}\;,
  \eqno(7.12)
  $$ 
we see that (7.11) reduces to the form taken by (7.9) when $h\rightarrow
h+1$. Moreover, for $h=1$ (7.9) becomes
  $$
  \de_v\biggl({j_0\over j_1}\,\de_ug_{N-k}\biggr)+\de_v\ln
  |j_1|\,{j_0\over j_1}\,\de_ug_{N-k}-g_{N-k}+{j_0\over
  j_1}\,\de_ug_{N-k-1}=0\;,
  \eqno(7.13)
  $$
which is just (7.8) for $i=N-k$, hence certainly true. This completes the
proof of (7.9).

In the proof that the substitution sequence terminates, we need (7.9) only
to derive the fact that, for $0\leq k\leq N$, 
  $$
  g_N={\cal D}_u^k\,g_{N-k}\;.
  \eqno(7.14)
  $$ 
The proof of (7.14) is also by induction, this time over $k$. Assuming
that the equation holds for some $k\leq N-1$, and writing (7.9) for the
particular case $h=k+1$, we have 
  $$
  \de_v\biggl({j_k\over j_{k+1}}\,\de_ug_N\biggr)+\de_v\ln |j_{k+1}|\,
  {j_k\over j_{k+1}}\,\de_ug_N-g_N+
  {\cal D}_u^{k+1}g_{N-k-1}=0\;.
  \eqno(7.15)
  $$
Now, (7.6) allows us to obtain
  $$
  g_N={\cal D}_u^{k+1}g_{N-k-1}\;,
  \eqno(7.16)
  $$
i.e., (7.14) for $k\rightarrow k+1$. Since for $k=0$ (7.14) is trivially
true (and even for $k=1$ it can be easily obtained by substituting (7.6)
into (7.8) for $i=N$), its validity for $0\leq k\leq N$ is established. 
In particular, we shall be interested in the case $k=N$, for which
  $$
  g_N={\cal D}_u^Ng_0\;. \eqno(7.17)
  $$

The equation analogous to (7.9) with $k=N$ is found by repeated
differentiation of (7.7) with respect to $u$; we have, for $i\ge1$,
  $$
  \de_v{\cal D}_u^i\,g_0+\de_v\ln |j_i|\,{\cal D}_u^i\,g_0-
  {\cal D}_u^{i-1}\,g_0=0\;,
  \eqno(7.18)
  $$
whose proof is perfectly analogous to that of (7.9). For $i=N$, (7.18)
becomes, using (7.17), 
  $$
  \de_vg_N+\de_v\ln|j_N|\,g_N-{\cal D}_u^{N-1}g_0=0\;.
  \eqno(7.19)
  $$

The final step in the proof consists in taking a further derivative with
respect to $u$ of (7.19), which gives, by (3.11), (7.6), and (7.17), 
  $$
  {j_{N+1}\over j_N}\,g_N=0\;,
  \eqno(7.20)
  $$
i.e., $j_{N+1}=0$. We have thus shown that the substitution
sequence of a PW$\!_N$ wave equation is upper terminating in $N$ steps. 
To prove that it is also lower terminating in $N$ steps, we could
manipulate (7.1)--(7.3) to show that $j_{-N-1}=\infty$, i.e.,
$l_{-N-1}=0$ by (3.17). However, it is much easier to notice that,
starting with the $u$-normal form of the wave equation, we can repeat the
proof above in a completely symmetric way to get directly $l_{-N-1}=0$. 
Hence, PW$\!_N\Rightarrow$ KN$\!_N$ and, since we know already that the
converse is also true, we can conclude that PW$\!_N$ and KN$\!_N$ are
equivalent properties.


\goodbreak
\section{8.}{Conclusions and open questions}

\noindent The results presented in this paper clarify the relationships
between several properties of wave equations related to the absence of
tails in their solutions, some of which were obvious, while some others
were never explicitly analysed in the literature, at least to our
knowledge. We have seen that the Huygens principle (HP) and the tail-free
property (TF) are satisfied by the same equations, that the characteristic
propagation property (CPP) is more general only in that it is satisfied, in
addition, by special two-dimensional equations, and that the
progressing wave propagation property (PW$\!_N$) and the solvability by the
Kundt-Newman method (KN$_N$) for two-dimensional wave equations are
equivalent. The two latter properties were not defined in more than two
dimensions, and they are not given in geometrical terms, despite the
motivation we gave for introducing PW$\!_N$ in section 3. However, we also
showed that in two dimensions PW$\!_0$ is equivalent to CPP, and
acquires thus a geometrical meaning. It would be therefore interesting to
investigate possible geometrical aspects of PW$\!_N$ equations for higher
$N$, and whether the definition of PW$\!_N$ can be meaningfully extended to
higher dimensions; we will comment here on these issues.

The result of section 6, that in $1+1$ dimensions CPP is equivalent to
PW$\!_0$, has the obvious corollary that no PW$\!_N$ equation with $N>0$
can satisfy CPP. This consequence is at first surprising, because from the
general form (3.3) of a progressing wave, one is tempted to think that, in
two dimensions at least, {\it all\/} PW$\!_N$ wave equations satisfy the
CPP. For, by choosing $R(u)$ and $S(v)$ to be of compact support, their
derivatives will also be of compact support, and all of $\phi$ will be made
of pieces sandwiched between null coordinate lines, so it will have no
tails. The problem with this argument is that, although solutions with no
tails {\it are\/} indeed obtained when $R$ and $S$ are of compact support,
the latter functions are not themselves the null data which are involved in
the definition of CPP; rather, the data are
  $$\eqalignno{
  &\varphi(u) = \sum_{i=0}^N f_i(u,0)\,R^{(i)}(u)
  + \sum_{i=0}^N g_i(u,0)\,S^{(i)}(0)\;, &(8.1)\cr
  &\psi(v) = \sum_{i=0}^N f_i(0,v)\,R^{(i)}(0)
  + \sum_{i=0}^N g_i(0,v)\,S^{(i)}(v)\;, &(8.2)}
  $$ and the $R(u)$ and $S(v)$ that correspond to generic $\varphi$ and
$\psi$ of compact support, will in general involve integrals of $\varphi$
and $\psi$ if $N>0$, and will thus be of non-compact support.

We can clarify this point by considering a typical example of PW$\!_N$
equations, obtained by separating the angular variables in the massless
Klein-Gordon equation
$\lapl\Phi=0$ in 4-dimensional Minkowski spacetime. Expanding $\Phi$ in
spherical harmonics as $\Phi(x) = \sum_{lm} \chi_{lm}(t,r)\, Y_{lm}
(\theta,\varphi)$, and rescaling the radial components to $\phi_{lm}:=
r\,\chi_{lm}$, we have that the $\phi_{lm}$ satisfy the $l$-dependent
equation
  $$
  \Biggl[ \de^2_t-\de^2_r+{l\,(l+1)\over r^2} \Biggr]\phi_{lm}
  \equiv \Biggl[ \de^2_{uv}+{l\,(l+1)\over (v-u)^2} \Biggr]\phi_{lm} = 0\;,
  \eqno(8.3)
  $$
where the null coordinates $u$ and $v$ are defined as in (2.5) and (2.6),
but now with $x$ replaced by $r$; the two-dimensional equation (8.3) is
PW$\!_l$ and can be solved exactly [17,20]. Since (5.23) is a necessary
condition for CPP, and in the case of (8.3) we have $U=V=0$, it follows that
CPP holds only if $W=0$; this is clearly false for $l\geq 1$. Let us see
explicitly the failure of CPP in the simple case $l=1$. Equation (8.3)
becomes
  $$
  \Biggl[ \de^2_{uv}+{2\over (v-u)^2} \Biggr]\phi(u,v) = 0\;, \eqno(8.4)
  $$
which is PW$\!_1$; its general solution is
  $$
  \phi(u,v) = {2\,R(u)\over v-u} + R'(u)
  - {2\,S(v)\over v-u} + S'(v)\;. \eqno(8.5)
  $$
We now look for the particular solution generated by null data with support
at one point, e.g.,
  $$\eqalignno{
  &\varphi(u) = \delta(u-u_0)\;, &(8.6)\cr
  &\psi(v) = 0\;. &(8.7)}
  $$
These data correspond to $R(u)$ and $S(v)$ satisfying the differential
equations
  $$
  R'(u) - {2\over u}\,R(u) + {2\over u}\,S(0)+S'(0)
  = \delta(u-u_0) \eqno(8.8)
  $$
and
  $$
  S'(v) - {2\over v}\,S(v) + {2\over v}\,R(0)+R'(0)=0\;, \eqno(8.9)
  $$
whose general solutions are, respectively,
  $$
  R(u) = {u^2\over u_0^2}\,\theta(u-u_0)+a\,u^2+S'(0)\,u+S(0)
  \eqno(8.10)
  $$
and
  $$
  S(v) =b\,v^2+R'(0)\,v+R(0)\;,
  \eqno(8.11)
  $$
with $a$ and $b$ arbitrary constants. Substituting (8.10) and (8.11)
into (8.5), and using the relations $R(0)=S(0)$ and $R'(0)=S'(0)$ that
follow, e.g., from (8.11), we have
  $$
  \phi(u,v)=\delta(u-u_0)+{2\over u_0^2}\,{uv\over v-u}\,
  \Bigl[\theta(u-u_0)+c\Bigr]\;,
  \eqno(8.12)
  $$
where $c:=(a-b)\,u_0^2$ can be prescribed arbitrarily. It is clear
from this expression that the only line $v=$ const on which $\phi$ is of
compact support is $v=0$; the choice $c=0$ gives a retarded wave, $c=-1$ an
advanced one, and any other value gives a solution which is non-zero
almost everywhere on $\cal M$!

Turning now to the question of whether it is meaningful to define the
class of wave equations whose general solution is a finite sum of
progressing waves, in dimension $m>2$, the simplest candidate for such an
equation would again be the example $\lapl\Phi=0$ in 4-dimensional
Minkowski spacetime, which satisfies HP. Using the expansion in spherical
harmonics given above, and (8.3), whose general solution is of the type (3.3)
with $N=l$ and all the functions carrying labels $l$ and $m$, we can write
  $$
  \Phi(x) = \sum_{l=0}^{+\infty} \sum_{i=0}^l \sum_{m=-l}^l
  \biggl(r\,f_{ilm}(t,r)\,Y_{lm}(\theta,\varphi)\,R_{lm}^{(i)}(t-r)+
  r\,g_{ilm}(t,r)\,Y_{lm}(\theta,\varphi)\,S_{lm}^{(i)}(t+r)\biggr),
  \eqno(8.13)
  $$
which contains derivatives up to arbitrarily high order of an
infinite number of free functions $R_{lm}$ and $S_{lm}$, and is
therefore not of the form (3.3). This shows that, at least in four
dimensions, the definition of PW$\!_N$ as it stands is empty for the case of
spherical wavefronts, and makes it plausible that it is ill-posed in
general. 

It is still useful to define a ``reduced'' PW$\!_N$ property for an
$m$-dimensional equation, in the sense that all two-dimensional equations,
obtained by separating out appropriately chosen coordinates, may have
progressing wave general solutions, as in the case of (8.3). A more
general example, in four dimensions, would be obtained with equations of the
type
		$$
  \Bigl[f(y,z)\,K_{t,x}+g(t,x)\,H_{y,z}\Bigr]\phi(t,x,y,z)=0\;,
  \eqno(8.14)
  $$
where $f$ and $g$ are nonvanishing functions and $K_{t,x}$, $H_{y,z}$
are suitable differential operators acting on the variables used as
subscripts. With the separation of variables $\phi(t,x,y,z) = \psi(t,x)\,
\varphi(y,z)$, we see that (8.14) satisfies the reduced progressing
wave propagation property if, for each value of $\alpha$ admitted by the
equation
		$$
  H_{y,z}\,\varphi(y,z)+\alpha\,f(y,z)\,\varphi(y,z)=0\;,
  \eqno(8.15)
  $$
the reduced wave equation		
  $$
  K_{t,x}\,\psi(t,x)-\alpha\,g(t,x)\,\psi(t,x)=0
  \eqno(8.16)
  $$
is PW$\!_N$ for some $N(\alpha)$. The high degree of dependence of the
reduction procedure---hence of the properties of the reduced
equation---from specific, noncovariant structure is evident. It is
thus very difficult to make general statements about this property and
its possible relationships with other ones we have discussed in this
paper, independently of the explicit form of the wave equation and the
choice of coordinates singled out for separation (i.e., of wave front).

It would be interesting to know whether a meaningful, truly $m$-dimensional
generalisation of the PW$\!_N$ property can be given, for another reason
as well. The expressions for the advanced or retarded Green function of
$\lapl\Phi=0$ in $m$-dim\-ens\-ional Minkowski space, which for even
$m\ge4$ are of the form [5,7]
  $$
  G_m(t',{\bf x}';t,{\bf x}) = \sum_{i=0}^N
  c_{mi}(|{\bf x}'-{\bf x}|)\,\delta^{(i)}(t'-t\pm |{\bf x}'-{\bf x}|)\;,
  \eqno(8.17)
  $$
with $N=(m-4)/2$, remind one of definition (3.2), and suggest a
possible connection between HP and progressing waves, this time of order
$N>0$, in dimension $m>2$. So far we have not been able to find such a
relationship, however. For now, PW$\!_N$ equations are left with no simple
interpretation in terms of tails, and no true generalisation to higher
dimensions, but their usefulness derives from the fact that they are
exactly solvable by the method of Kundt and Newman [20], as we saw in
section 7, and from their relationship to Toda lattices [25].

Taking into account all of these remarks, there is little doubt that, as
far as studies of tails are concerned, the definition that should be
preferred is, for its generality, the one based on HP or, equivalently,
on TF. With a notion of wave tails that is now completely clear and
unambiguous, one can conduct further research in several directions. 
Here is a list of some problems that one may address:

\item{(i)} Which are the conditions on $g^{ab}$, $H^a$, and $K$ that make
(1.3) tail-free?  Because of the geometrical interpretation of $g_{ab}$
as a metric in $\cal M$, it is natural to restate this question by asking
in which spacetimes (i.e., for which class of $g^{ab}$) is the wave equation
tail-free. This is sometimes called the {\it Hadamard problem}, and has
received some attention over the years [6,26,27].

\item{(ii)} What happens for vector and tensor fields?  These cases include
the propagation of electromagnetic and gravitational waves [15,26,28] and are
more realistic---though somewhat more complicated---than the scalar field one.

\item{(iii)} Which are the physical consequences and effects of wave
tails?  The situation that has been investigated most in detail is that of
radiation from compact objects [2,3,8]. The corresponding problem in a
cosmological background [11] has apparently been neglected, probably
because of the belief that any observable effect should be extremely small.
However, in this case curvature never drops off and scattering continues
forever, so it might convey a relevant amount of radiation in the interior of
the light cone [9]. This brings immediately out a new problem, namely
{\it how much\/} radiation goes into the tail, which leads us to the next
question.

\item{(iv)}  How to quantify tails?  A very natural and intuitive way
would be to calculate a reflection coefficient; this is possible when
backscattering is localised and there are regions in which the field is
free (even asymptotically, or after a suitable coordinates transformation
has been performed [8]). In a cosmological context, however, the property
that makes the phenomenon potentially interesting---i.e., the fact that
backscattering takes place always and everywhere---at the same time prevents
us from defining purely ingoing and outgoing solutions of the wave equation,
and hence from computing reflection and transmission coefficients [9]. It is
necessary to find alternative ways to quantify tails, perhaps based on the
ratio between their energy content and the total energy of the field.


\goodbreak
\section{Acknowledgements}{} 

\noindent This work was supported by the Commission of the European
Communities under the DG-XII contract no.\ CI1*-0540-M(TT) and the DG-III
contract no.\ ECRU002, and by the Instituts Internationaux de Physique et
de Chimie Solvay.



\section{References}{}\frenchspacing \eightpoint
\parindent=15pt
\def\dash{---\hskip-1pt---}

\def\AJP{{\it Am. J. Phys. }}
\def\CQG{{\it Class. Quantum Grav. }}
\def\GRG{{\it Gen. Rel. Grav. }}
\def\JMP{{\it J. Math. Phys. }}
\def\JPA{{\it J. Phys. A: Math. Gen. }}
\def\PCPS{{\it Proc. Cambridge Philos. Soc. }}
\def\PL{{\it Phys. Lett. }}
\def\PRD{{\it Phys. Rev.} D }

\item{[1]} Huygens C 1690 {\it Trait\'e de la lumi\`ere\/} (Leiden: Van
der Aa)

\item{[2]} Blanchet L and Damour T 1992 Hereditary effects in
gravitational radiation \PRD {\bf46} 4304--19;\hfill\break
Blanchet L and Sch\"afer G 1993 Gravitational wave tails
and binary star systems \CQG {\bf10} 2699--721 

\item{[3]} Wiseman A G 1993 Coalescing binary systems of compact
objects to (post)$^{5/2}$-Newtonian order. IV. The gravitational wave tail
\PRD {\bf48} 4757--70

\item{[4]} Hadamard J 1952 {\it Lectures on Cauchy's Problem in Linear
 Partial Differential Equations\/} (New York: Dover)

\item{[5]} Courant R and Hilbert D 1962 {\it
Methods of Mathematical Physics\/} (New York: Wiley) vol 2

\item{[6]} Friedlander F G 1947 Simple progressive solutions of the wave
equation \PCPS {\bf43} 360--73

\item{[7]} Soodak H and Tiersten M S 1993 Wakes and waves in $N$
 dimensions \AJP {\bf61} 395--401

\item{[8]} Price R H 1972 Nonspherical perturbations of relativistic
gravitational collapse. I. Scalar and gravitational perturbations
\PRD {\bf5} 2419--38;\hfill\break
Misner C W, Thorne K S and Wheeler J A 1973 {\it Gravitation} (San Francisco:
Freeman)

\item{[9]} Faraoni V and Sonego S 1992 On the tail problem in cosmology
\PL {\bf 170A} 413--20

\item{[10]} Visser M 1993 Acoustic propagation in fluids: an
 unexpected example of Lorentzian geometry (pre\-print gr-qc/9311028)

\item{[11]} Ellis G F R and Sciama D W 1972 Global and non-global
problems in cosmology, {\it General Relativity, Papers in Honour of J.
L. Synge} ed L O'Raifeartaigh (Oxford: Clarendon) 35--59

\item{[12]} DeWitt B S and Brehme R W 1960 Radiation damping in a
 gravitational field {\it Ann. Phys. (N.Y.)\/} {\bf 9} 220--59

\item{[13]} Fox R et al 1969 Do faster-than-light group velocities
imply violation of causality? {\it Nature\/} {\bf 223} 597;\hfil\break
\dash\ 1970 Faster-than-light group velocities and causality violation {\it
Proc. R. Soc.\/} A {\bf 316}, 515--24;\hfil\break
Bers A et al 1971 The impossibility of free tachyons {\it Relativity
and Gravitation\/} C G Kuper and A Peres (New York: Gordon and Breach)

\item{[14]} Wald R M 1984 {\it General Relativity\/} (Chicago: University
 of Chicago Press)

\item{[15]} Sonego S and Faraoni V 1992 Huygens' principle and
 characteristic propagation property for waves in curved space-times \JMP
 {\bf33} 625--32

\item{[16]} Friedlander F G 1975 {\it The Wave Equation on a Curved
 Space-time\/} (Cambridge: Cambridge University Press)

\item{[17]} Couch W E and Torrence R J 1986 A class of wave equations
 with progressive wave solutions of finite order \PL {\bf 117A} 270--4

\item{[18]} Ward R S 1987 Progressing waves in flat spacetime and in
plane-wave solutions \CQG {\bf 4} 775-8

\item{[19]} Torrence R J and Couch W E 1988 Progressing waves on
 spherical spacetimes \GRG {\bf20} 343--58

\item{[20]} Kundt W and Newman E T 1968 Hyperbolic differential equations
 in two dimensions \JMP {\bf9} 2193--210

\item{[21]} Torrence R J and Couch W E 1985 Transparency of de Sitter
 and anti-de Sitter spacetimes to multipole fields \CQG {\bf2} 545--53

\item{[22]} Bombelli L, Couch W E and Torrence R J 1991 Wake-free waves
 in one and three dimensions \JMP {\bf 32} 106--8

\item{[23]} Richtmyer R D 1981 {\it Principles of Advanced Mathematical
 Physics\/} (New York: Spring\-er-Verlag) vol I, p 51

\item{[24]} Torrence R J 1990 Self-adjoint acoustic equations with
progressing wave solutions \JPA {\bf23} 4107--15

\item{[25]} Torrence R J 1987 Linear wave equations as motions on a
Toda lattice \JPA {\bf20} 91--102;\hfill\break
Bombelli L, Couch W E and Torrence R J 1992 Solvable
systems of wave equations and non-Abelian Toda lattices \JPA {\bf 25}
1309--27

\item{[26]} K\"unzle H P 1968 Maxwell fields satisfying Huygens's principle
{\it Proc. Camb. Phil. Soc.} {\bf 64} 779--85;\hfill\break
Carminati J and McLenaghan R G 1986 An explicit
determination of the Petrov type $N$ space-times on which the
conformally invariant scalar wave equation satisfies Huygens' principle
{\it Ann. Inst. Henri Poincar\'e, Phys. Th\'eor.} {\bf44}
115--53\hfill\break
\dash\ 1987 An explicit determination of the spacetimes on
which the conformally invariant scalar wave equation satisfies Huygens'
principle. Part II: Petrov type D space-times {\it Ann. Inst. Henri
Poincar\'e, Phys. Th\'eor.} {\bf47} 337--54\hfill\break
\dash\ 1988 An explicit determination of the spacetimes on which the
conformally invariant scalar wave equation satisfies Huygens' principle.
Part III: Petrov type III space-times {\it Ann. Inst. Henri Poincar\'e,
Phys. Th\'eor.} {\bf48} 77--96

\item{[27]} McLenaghan R G 1969 An explicit determination of the empty
space-times on which the wave equation satisfies Huygens' principle {\it
Proc. Camb. Phil. Soc.} {\bf 65} 139--55

\item{[28]} Noonan T W 1989a Huygens's principle for the electromagnetic
vector potential in Riemannian spacetimes {\it Ap. J.} {\bf341} 786--95
\hfill\break
\dash\ 1989b Huygens's principle for the wave equation for second-rank
tensor fields {\it Ap. J.} {\bf343} 849--52;\hfill\break
W\"unsch V 1990 Cauchy's problem and Huygens' principle
for the linearized Einstein field equations {\it Gen. Rel. Grav.} {\bf22}
843--62;\hfill\break
Caldwell R R 1993 Green's functions for gravitational
waves in FRW spacetimes \PRD {\bf48} 4688--92

\vfill\eject
\end